\documentclass{elsart}

\usepackage{amssymb,amsmath}
\usepackage{times}  
\usepackage{graphicx}
\usepackage{theorem}
\usepackage[latin1]{inputenc}
\usepackage{latexsym}
\usepackage{url}

\theoremstyle{break}
\newtheorem{example}{Example}
\theorembodyfont{\itshape}
\theoremheaderfont{\scshape}

\begin{document}

\begin{frontmatter}

\title{Comparison and validation of community structures in complex networks}

\author{Mika Gustafsson,  Michael H{\"o}rnquist$^*$,  Anna Lombardi }
\corauth[cor1]{Corresponding author}
\address{Division of Physics and Electronics, Department of Science
  and Technology, Linköping University, SE-601 74 Norrköping, Sweden}
\ead{[mikgu,micho,annlo]@itn.liu.se}

\begin{abstract}
The issue of partitioning a network into communities has attracted a great
deal of attention recently.
Most authors seem to equate this issue with the one of finding the
maximum value of the modularity, as defined by Newman.
Since the problem formulated this way is NP-hard, most effort
has gone into the construction of search algorithms, and less to
the question of other measures of community structures, similarities
between various partitionings and the validation with respect to
external information.

Here we concentrate on a class of computer generated networks and on
three well-studied real networks which constitute
a bench-mark for network studies; the karate club, the US college
football teams and a gene network of yeast.
We utilize some standard ways of clustering data (originally
not designed for finding community structures in networks) and show
that these classical methods sometimes outperform the newer ones.
We discuss various measures of the strength of the modular structure,
and show by examples features and drawbacks.
Further, we compare different partitions by applying some
graph-theoretic concepts of distance, which indicate that one of the
quality measures of the degree of modularity corresponds quite well
with the distance from the true partition.
Finally, we introduce a way to validate the partitionings with respect
to external data when the nodes are classified but the network
structure is unknown.
This is here possible since we know everything of the computer
generated networks, as well as the historical answer to how the karate
club and the football teams are partitioned in reality.
The partitioning of the gene network is validated by use of the
Gene Ontology database,
where we show that a community in general corresponds to a biological process.

\end{abstract}

\end{frontmatter}

\section{Introduction}

Complex networks, i.e., assemblies of nodes and edges with nontrivial
properties, can be used to describe systems in many
different fields, such as sociological (scientific collaborations and
structure of organizations, 
biological (proteins and genes interactions 
and technological (Internet and the web).
These systems are composed of a large number of
interacting agents, and the complexity  originate partly
from the heterogeneity in their interaction patterns. Given this high
degree of complexity, it is often necessary to divide a network into
different subgroups to facilitate the understanding of the
relationship among different components
\cite{girvan02,zhou2003b,ravasz2002}.
Outlines of recent work are
given in, e.g.,  \cite{evans2004,Danon-05} together with broad
discussions of relevant literature.

\begin{figure}[h]
  \centering
      \includegraphics[width=0.5\linewidth]{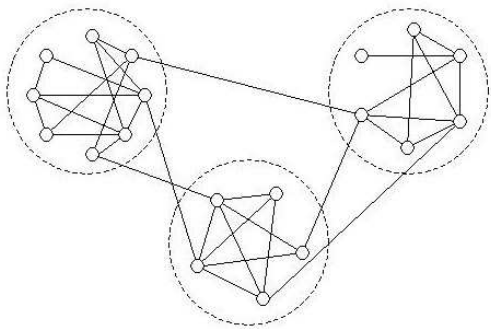}
   \caption{Example of a small network with community structure.}
  \label{fig:communities}
\end{figure}

In recent years there has been an increasing interest in the
properties of networks, and the property of community
structure has attracted great attention. The vertices in the networks are
often found to cluster into tightly-knit groups with a high density of
within-group edges and a lower density of between-group edges.
Clustering techniques have  acquired a dominant role
among the tools used to decompose the network into functional units.
Community structure is a topological property of networks and it is
linked to the concept of classification of objects in categories.
The working definition of community is general, but
ambiguous.
There is no generally accepted formal definition,
but an informal one is
``a subset of nodes within the graph such that
connections between the nodes are denser than connections with the
rest of the network'', see Fig.\ref{fig:communities}.
That is,  a community can be seen,
depending on the context, as a class, group, cluster etc.
Communities in a social network might represent real social groupings,
perhaps by interest or background; communities in a citation network
might represent related persons on a single topic, communities in a
gene networks might represent specific biological processes,
communities on the web might represent pages  on related
topics, and so on.
Being able to identify these communities could help us to
understand and exploit these networks more efficiently.

Since the issue of community detection has acquired 
relevance in several fields, a great interest has
raised on the algorithms used to determine communities in a network.
Several algorithms to determine communities
are known in the literature.
A survey
of the different approaches can be found in \cite{girvan04,newman04},
and especially in \cite{Danon-05} where these approaches are evaluated
with respect to computational cost and sensitivity.
In the present paper, we return to two classical methods of finding
groups among data---the hierarchical clustering and the K-means.
Both of these methods are well-known within the applied statistics
community \cite{Speed}, but at least the former has in some
papers been discarded
as a less suitable way of finding communities in a network
\cite{girvan02,newman04}.
One of our  results is that this rejection was a bit
premature---with a suitable metric for the distances between the
nodes,   the result of such a hierarchical clustering
can outperform some of the more modern approaches.
The K-means has not, to the best of our knowledge, been utilized
before for finding community structures in complex networks.

The present paper focuses on networks with a single type of vertex and a
single type of undirected, unweighted edges.
As examples, we consider simulated networks (originally described
in \cite{girvan02}) and three real examples from the
literature---Zachary's karate club \cite{karate}, the college football
teams in US division one for the year 2000 \cite{girvan02}
and a gene network of the yeast
\emph{Saccharomyces cervisae} \cite{milo2002}.
By using different quality measures, we explore how well the two
methods we study perform, and by using distance measures for
partitions
we also investigate how similar the communities found by different
algorithms are.
Finally, we introduce an external validation method which works
also for only partially known networks.
The article is written to be as self-contained as possible, which means that we
introduce rather carefully also concepts that are well described
elsewhere in the literature.
However, several of the sources are scattered among different
scientific areas, and therefore we consider it as meaningful to repeat
them here. Careful references are always given, though, when the
concepts are not new.

The idea of the disposition of this article is to carefully
introduce new concepts
in almost direct relation to the networks to which they are applied
for the first time. This takes the following form:
\begin{itemize}
   \item Section \ref{section:clustering} introduces the tools we
     utilize for finding community structures in the networks, particularly
     \begin{itemize}
       \item metrics for distances between the nodes.
       \item clustering algorithms (hierarchical clustering and
         K-means).
     \end{itemize}
   \item Section~\ref{section:karate},  the karate club network, with
     the concepts
     \begin{itemize}
       \item modularity.
       \item Silhouette index.
       \item a null hypothesis by rewiring.
       \item measures and indices of the similarity of two partitionings.
     \end{itemize}
   \item Section~\ref{section:CG},  computer generated networks,
     with illustrations of most of the concepts introduced thus far.
   \item Section ~\ref{section:football}, US college football teams, with
     \begin{itemize}
       \item a novel coherence score, measuring how well the detected modules
         correspond to an external classification of the units of the
         network.
     \end{itemize}
   \item Section~\ref{section:gene_network}, a gene network from yeast.
     This time, no ``true'' partitioning exists, and we apply the
     different methods introduced earlier in this article in order
     to see whether we can obtain a biological meaningful division
     of the network.
   \item Section~\ref{section:discussionandconclusions},  a
     general discussion of the results obtained and conclusion of the
     paper
\end{itemize}

\section{Detecting  community structures}
\label{section:clustering}

This section introduces the algorithms we utilize for detecting
communities.
It comprises several ideas scattered through the
literature, and its purpose is to make the article self-contained.
However,  the reader is still referred to the references for more
detailed descriptions of the central concepts.

\subsection{Distance between nodes}

First we need a way to measure distance between nodes in the networks.
The most common way is to consider the geodesic, that is, the shortest
path (counted in number of links) connecting two vertices. The
geodesic distance between two nodes is then just the minimum number of
links
which separate them.
For future reference we denote the matrix of all pairwise
geodetic distances as $G$.

In \cite{girvan02} Girvan and Newman propose to
define the distance between vertices as the total number of paths that run
between them.
However, the number of paths between any two vertices is infinite (unless it is
zero) so  paths of length $\ell$ are weighted with a factor
$\alpha^{\ell}$ with $\alpha$ small, so that the weighted count of the
number of paths converges. In this way, long paths contribute
with less weight than those that are short. If $A$ is the
adjacency matrix of the network, such that $A_{ij}$ is unity if there is
an edge between vertices $i$ and $j$ and zero otherwise, then the
distances are given by the elements of the matrix $W$, 
calculated as
\begin{equation}
W=\sum_{\ell =0}^{\infty}(\alpha A)^{\ell}= (I-\alpha A)^{-1}.
\label{sumpaths}
\end{equation}
For the sum to converge, $\alpha$ must be chosen smaller than the
reciprocal of the largest eigenvalue of $A$.
Both these definitions of distances give reasonable results for
community structures, but in some cases they are less successful.
Different authors have used different approaches for the choice of
the weights but it has not, to the best  knowledge of the present authors, been made a
systematic comparison between
clustering algorithms implemented with different choices of distances.

Both measures result in a matrix ($G$ or $W$) of dimension $n\times n$
(where $n$ denotes the 
number of nodes in the network) whose elements are the
distance between the nodes in the network.

Eventually, we utilize as distance measure between node $i$ and $j$
the euclidean distance between the $i$th and $j$th row in one of the
matrices ($G$ or $W$) above. That is, we take
\begin{equation}
\text{dist}(\text{node $i$},\text{node $j$}) = \sqrt{\sum_k
  (P_{ik}-P_{jk})^2},
\label{euclidean}
\end{equation}
where $P=G$ for the geodesic distance and $P=W$ for the sum
of all paths.
This way, the issue of finding communities in a network becomes
algorithmically identical to finding clusters from co-variation within
a series of experiment, and standard routines available in, e.g.,
MatLab, R, etc. can be used.
A similar approach with respect to (\ref{euclidean}) and
hierarchical clustering, but with a different $P$, has been
suggested in \cite{Rives-03} and applied to a protein-protein
interactions network.

\subsection{Community detecting algorithms}

The following two standard algorithms have been implemented and tested:
\begin{itemize}
\item Hierarchical clustering.
\item K-means algorithm.
\end{itemize}
Of course, our choice of tested algorithms is not intended to be
exhaustive, 
but we have focused on some algorithms that are well-known
clustering methods, but have not been applied systematically
to complex networks.

Hierarchical clustering \cite{fisher96} is an unsupervised procedure of transforming a
distance matrix, which is a result of pair-wise similarity measurement
between elements of a group, into a hierarchy of nested partitions.
It is an agglomerative procedure, which means that it starts
with as many clusters as there are nodes, 
i.e., 
each node forms a cluster containing only itself.
Iteratively the number of clusters is reduced by a merging of the two most
similar clusters until only one cluster remains.
Once several nodes have been linked together, a linkage rule is
needed to determine if two clusters are sufficiently similar to be
linked together.
Different linkage rules have been proposed. In this paper 
{\em complete linkage}, also called {\em furthest neighbour} is used. 
This method utilizes the largest distance between nodes in two groups.
Explicitly, it takes the form
\begin{equation}
d(r,s)=\max_{i,j} \left(  \mbox{dist}(x_{r_i}, x_{s_j}) \right),   
\begin{array}{l}
i=1,\cdots,n_r \\ 
j=1,\cdots,n_s
\end{array}
\end{equation}
where $n_r$ and $n_s$ are the numbers of nodes in clusters $r$ and $s$,
respectively, and
$x_{r_i}$ denotes the $i$th node in cluster $r$.

K-means \cite{huang97} is a clustering algorithm that is widely used
 when working with temporal data where
the elements should be grouped on the basis of their time
profile.
Data are clustered into $N$ 
mutually exclusive clusters, where $N$ has to be chosen
beforehand. 
It uses an iterative algorithm that minimises the sum of
distances from each object to its cluster centroid, over all
clusters. 
The algorithm moves the objects in a deterministic fashion between the 
clusters until the sum
cannot be decreased further. 
The result is a set of clusters that
is as compact and well-separated as possible, given the initial
partitioning. 
A drawback of the K-means algorithm is that it often converges to
a local optimum.
This problem can be somewhat remedied  by choosing multiple
starting points.
For what we are aware of there have not yet been systematic studies of the
use of K-means algorithm to detect community structure in networks.

\section{Zachary's karate club}
\label{section:karate}

Zachary observed 34 members of a karate club over a period of 2
years. During the course of the study, a disagreement developed
between the administrator of the club and the club's instructor, which
ultimately resulted in the instructor leaving and starting a new
club taking about half of the original club's members with him.

\begin{figure}[h]
  \centering
    \includegraphics[width=0.5\linewidth]{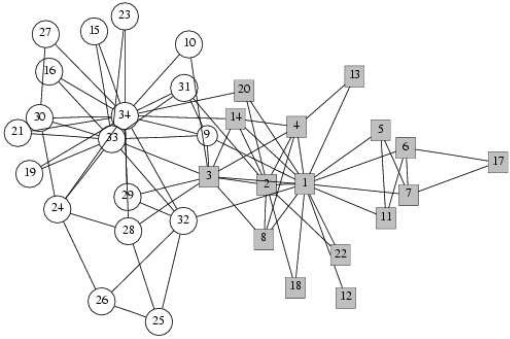}
   \caption{Zachary's friendship network of the karate club.}
  \label{fig:karate}
\end{figure}

Zachary constructed a network of friendship between members of the
club, using a variety of measures to estimate the strength  of ties
between individuals. Here we use a simple unweighted version of his
network with the attempt to identify the factions involved in the
split of the club (see \cite{girvan02}, \cite{karate}).
The network\footnote{The network can be downloaded from
  \url{http://vlado.fmf.uni-lj.si/pub/networks/data/UciNet/UciData.htm} and
  the graphical representation of the network is obtained from
  \url{http://www-personal.umich.edu/~mejn/networks/}; see \cite{girvan02}.}
consists of 34 nodes and 78 links as illustrated in
Fig.~\ref{fig:karate}; squares denote the supporters of the trainer
(node 1) and circles represent the supporters of the administrator
(node 34).

When we apply the community detecting algorithms described in
Sec.~\ref{section:clustering} above,
we also need a way of deciding upon the number
of communities we should have.
One way of getting this number, when we do not have prior
knowledge, is to measure the quality of the partitioning itself
and simply pick the number which results in the ``best'' clustering.
However, in accordance with the ambiguity mentioned in the
introduction about what is meant
by community structure, there exist several different
measures for the quality of the partitioning of a set of nodes
into communities.
Here we will discuss two such measures, the \emph{Silhouette index}
\cite{azuaje} and the \emph{modularity} \cite{girvan02}.

\subsection{Evaluation of the strength of the community structure, modularity}

The first measure we consider is Newman's
\emph{modularity} \cite{Newman-03}, which has
become a kind of de facto standard for measuring the quality of a
partitioning (although there are alternatives coming, e.g.,
\cite{Muff-05,Massen-05}).

The modularity is defined the following way:
Given a particular division of a network into $k$ communities, let $e$
denote a $k\times k$ matrix whose element $e_{ij}$ is the fraction of
all edges
in the network which connect
vertices in community $i$ to those in community $j$.
The trace of this matrix $\text{Tr} e= \sum_i e_{ii}$ gives the
fraction of edges in the network that connect vertices in the same
community. A good community division should have a high value of the
trace, but the trace on its own is not a good indicator of the quality
of the division since, for example, placing all vertices in a single
community would give the maximal value of $\text{Tr} e = 1$ without
giving any information about the community structure. The row (column)
sum is then defined as $a_i=\sum_j e_{ij}$ and it represents the
fraction of edges that connect to vertices in community $i$. In a
network where edges fall between vertices without regard for the
communities they belong to, it holds $e_{ij}=a_ia_j$.
The modularity is therefore defined as
\begin{equation}
Q=\sum_i \left( e_{ii}-a_i^2 \right). 
\label{modularity}
\end{equation}
It measures the fraction of edges in a community, minus the
expected value of the same quantity  in a network with the
same community divisions but random connections between the
nodes.
If a particular division gives no more within-community edges that
would be expected by random chance the modularity is
zero. Values other than 0 indicate deviations from randomness, and
as a rule of thumb, values above 0.3 indicate a modular structure
\cite{Newman-04}.
In practice, values above 0.7 are rare, and indicate a very
clear structure.
However, also Erd\"os-R\'enyi (ER) random graphs can possess a very
high modularity, as shown in \cite{Guimera-04}. The reason is that
there are so many different ways to partition a network, that it is
likely that there should be at least one partition where the intra-density of
links within a cluster greatly exceeds the one obtained by chance.
Because of this, we consider below in Section \ref{Rewiring} a null hypothesis
where the networks are rewired.

\begin{figure}[h]
  \centering
      \includegraphics[width=0.9\linewidth]{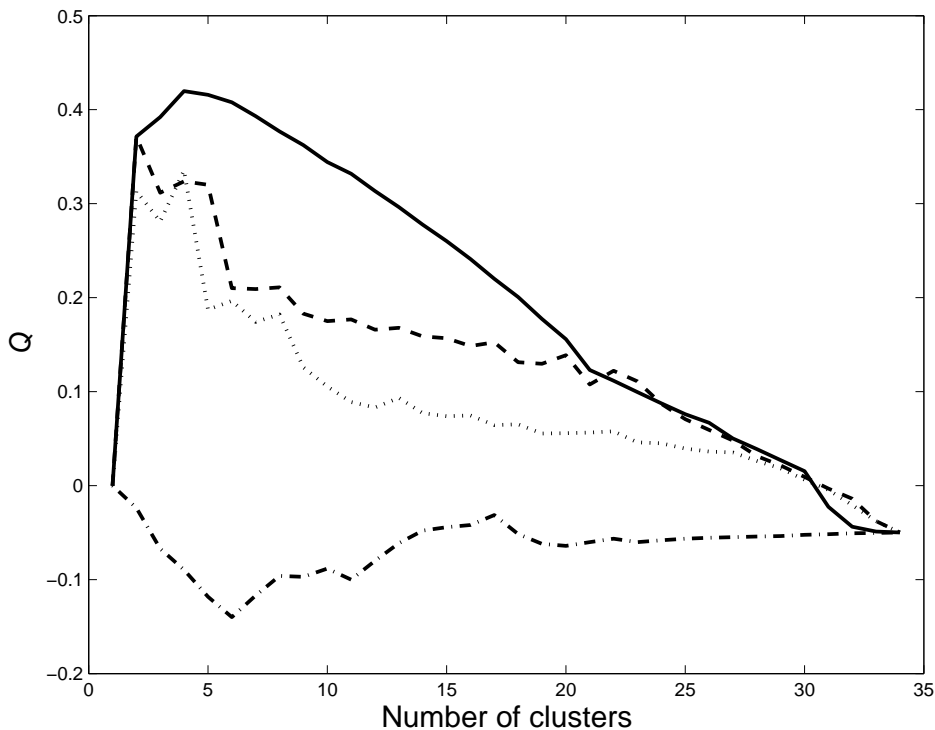}
  \caption{Modularity $Q$ for the communities detected in
    Zachary's karate club. \textit{Solid curve}: hierarchical
    clustering (shortest path), \textit{dashed curve}: 
    K-means clustering (sum of paths), \textit{dotted curve}:
    K-means clustering (shortest path), \textit{dash-dotted curve}:
    hierarchical clustering (sum of paths).}
  \label{fig:karate_plots}
\end{figure}

We apply the K-means and the hierarchical clustering methods to the
Karate club, both when the distances are the geodetics and when
they are a sum over all paths.
By calculating the modularity  for every
possible number of partitions, from one up to 34, we can easily see what the
``best'' division is.
For each partition the modularity $Q$ is computed
and is plotted in Fig.~\ref{fig:karate_plots} (for the K-means, we
took the division among the  repetitions with the same number of
communities resulting in the highest $Q$, and discarded the rest).
We can immediately see that the hierarchical
algorithm applied with the distance given by the sum of all the paths
(\ref{sumpaths}) gives always negative values of $Q$; this means that
the partitions obtained do not correspond to communities of the
network.
The other algorithms, instead, can all partition the network into
communities. For each algorithm we have analysed the partition
corresponding to the highest modularity $Q$, and the results are
presented in Table \ref{table:modularity_karate}.\footnote{We note that
our peak value, $Q=0.4198$, slightly exceeds the maximum we have
found in the literature, $Q=0.4188$, in \cite{Duch-05}.}
\begin{table}[h]
 \caption{Karate club: highest modularities obtained using different
   approaches and their corresponding community structure computed for
   the true network ($Q$) and for the rewired network representing the
   null hypothesis ($Q_{H_0}$).}
   \centering
   \begin{tabular}{|l||c|c||c|c|}
\hline
Algorithm  & $Q$  & No of clusters & $Q_{H_0}$ & No of clusters \\
\hline \hline
Hierarchical (shortest path)   & 0.42 & 4 & $0.22\pm 0.04$ & 9 \\
K-means, (sum of paths)         & 0.37 & 2 & $0.15\pm 0.02$ & 9\\
K-means, (shortest path)        & 0.34 & 4 & $0.18\pm 0.07$ & 3\\
\hline
    \end{tabular}
   \label{table:modularity_karate}
  \end{table}

The K-means
algorithm used with the sum of paths distance between nodes detects
two communities and all the nodes are partitioned according to the
sociological division that took place in the club (circles and squares
of Fig.~\ref{fig:karate} are divided into two separate groups).
When both the K-means algorithm and hierarchical clustering are used
with the distance given by the shortest path, they detect 4
communities. Looking in details  we see that in both cases the 4
communities represent a further division of the 2 factions created in
the club.
This means that the union of communities 1 and 2 is exactly
the group of supporters of the trainer and the union of communities 3
and 4 contains all the supporters of the administrator.

\subsection{Evaluation of the strength of the community structure,
  Silhouette index}

The second measure we discuss is the \emph{Silhouette index}
\cite{silhouette,azuaje,Bolshakova-03}.
This measure is wide-spread in the context of
clustering based on co-variation over several experiments, and
since we explore such methods here, a discussion of the measure
is most appropriate.

For each cluster, one can calculate the Silhouette index, $S_j$, which
characterizes the heterogeneity and isolation properties of the cluster.
The \emph{Global Silhouette index},
$GS$, is  the mean of all the Silhoutte indices (one for each
cluster) for the set, and can be used as an effective validity index.

A drawback of the Silhouette index is that a community consisting of
only one node is considered to be a perfect partitioning, i.e., the
confidence indicator becomes unity.
Thus, this measure will be inclined towards such clusters, which
makes it a less suitable candidate for measuring the quality of a
partitioning.
This effect can be eliminated by modifying the
average by discarding  all the terms in the sum
corresponding to clusters with only one element.
This is what has been implemented here.

\begin{figure*}[h]
  \begin{tabular}{cc}
      \includegraphics[width=0.45\linewidth]{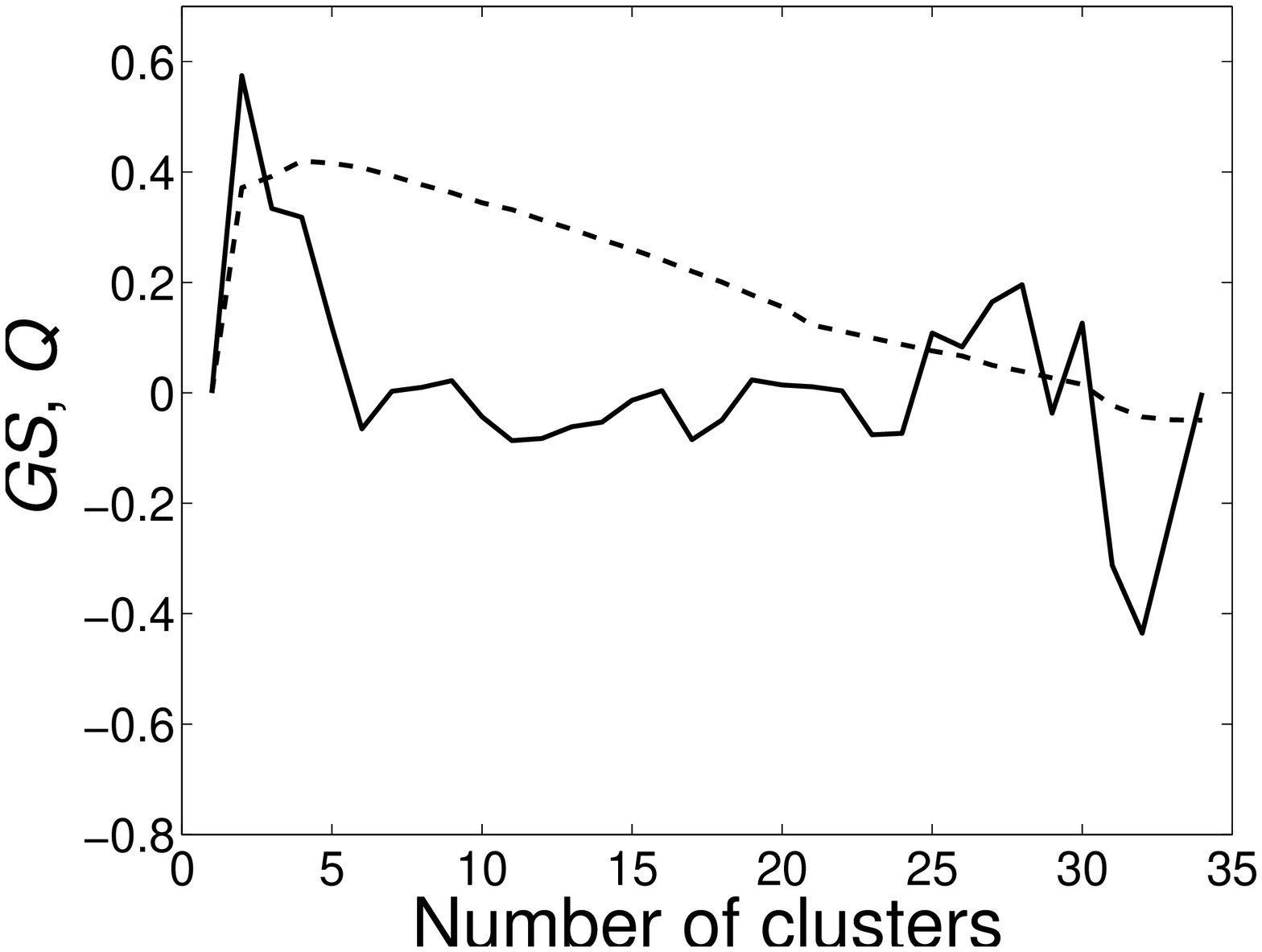}
      &
      \includegraphics[width=0.45\linewidth]{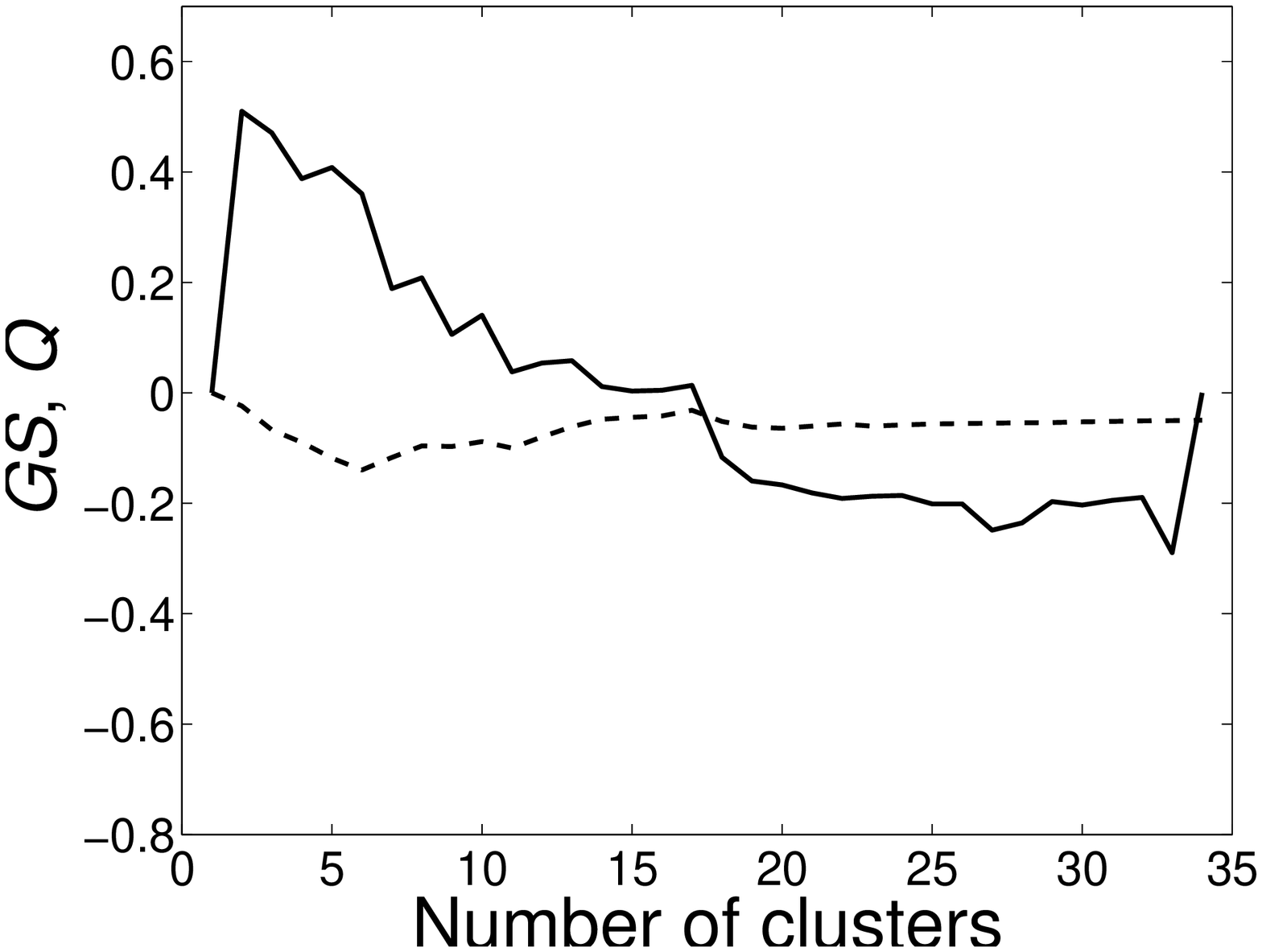}
    \end{tabular}    
  \caption{Global silhouette index $GS$ (\textit{solid curve}) for
    the communities
    detected in Zachary's karate club with hierarchical
    clustering  (\textit{left}: shortest path, \textit{right}: sum
    of paths). The modularities  $Q$
    (\textit{dashed curve}) are also shown as comparisons.}
  \label{fig:karate_silhouette}
\end{figure*}

This index has been computed for all the partitions detected by the
hierarchical algorithm with both the shortest path distance and the
sum of paths distance between nodes, i.e., for one partitioning which
works reasonable and for one which gives nonsensical results for the
modularity.
The results are shown in Fig.~\ref{fig:karate_silhouette}.
The figure shows that in both cases the
Silhouette index has the maximum value for $N=2$, that is, when the set is
divided into two clusters.
Checking the corresponding partitions we see that when the
hierarchical clustering (shortest path) is used, the maximum of the Global
Silhouette index really corresponds to the real structure of the set.
On the other hand the partition corresponding to two communities
detected by hierarchical clustering (sum of paths) has one
cluster containing only one node (node 34: the administrator) and all the
other nodes are grouped together. This partition is clearly quite far
from the original one while the Global Silhouette index still
has a numerical value
that in literature is considered to indicate a community structure.

The Global Silhouette index is obviously less suited for estimating the
quality of a network partitioning, and in the sequel of the
present paper we will only utilize the modularity for measuring the
degree of community structure.
Further, since the algorithm based on hierarchical clustering (sum of
paths)  does not seem to work,
we will also exclude that one from all further considerations.

\subsection{Rewiring}
\label{Rewiring}

The results obtained are here tested with respect to the null
hypothesis that the found community structure is solely due to the
degree distribution.
This testing is necessary, since it has been shown that also random
networks can possess a high modularity \cite{Guimera-04}.
Each network is rewired according to the algorithm presented in
\cite{maslov2002}. This algorithm randomly rewires a network
while preserving the degrees of each node.
In this way, we investigate if the communities detected by the
algorithms might have occurred by the degree distribution only,
or if they  represent an intrinsic structure of the network beyond the
most obvious one.
The three different
clustering schemes (hierarchical clustering (sum of paths) is
discarded) are applied to the rewired network and the
modularity $Q$ is computed. This procedure is
repeated 100 times, and the results are shown in Table
\ref{table:modularity_karate}.
For each clustering
procedure the mean of the $Q$ value is computed over all the
repetitions. The  $Q$ value is reported together with the
corresponding standard deviation. The values of the modularity
 obtained for the rewired networks are much smaller with
respect to the true network; this shows that the partition detected in
the network have not occurred from the degree distribution only.

\subsection{Distance between partitions}

A less studied property, at least in the recent physics literature
on networks, is how to compare two different partitionings of
the same network with each other.
Different definitions of distance between partitions can be found in
the literature (see, e.g., \cite{Milligan-81} for an early comparison
of different measures), and here we discuss  two of these.
In the next subsection, we will compare these measures with two
indices that have recently been advocated within this context.
For both measures, we need the concept of \emph{meet} of two
partitionings.
Let $A$ and $B$ denote two partitionings of the whole set, with
corresponding elements $a_i$, $i=1,\cdots,|A|$ and $b_i$,
$i=1,\cdots,|B|$, which are sets themselves.
The meet is then the set $C$ given by
\begin{equation}
  C=\bigcup_{i=1}^{|A|}\bigcup_{j=1}^{|B|} \left\{a_i \bigcap b_j\right\}.
  \label{meet}
\end{equation}
The two distance measures we discuss here are:
\begin{itemize}
  \item[$m_{\text{moved}}$:] 
    The distance between the partitioning $A$ and the meet $C$, given
    the partitioning $B$, is defined as  the minimum number of
    elements that must be \emph{moved} between the partitions so that $A$ and
    $C$ become identical \cite{gusfield2002}. The distance between $A$
    $B$, also denoted as the \emph{total} distance, is then obtained
    as the sum of the distance between $A$ and
    $C$ and the distance between $B$ and $C$. (Alternative, but
    equivalent, definitions can be found in \cite{gusfield2002,vandongen}.)
    
  \item[$m_{\text{div}}$:] The distance between the partitioning $A$ and the meet $C$, given
    the partitioning $B$, is defined as the minimum number of 
    \emph{divisions}  that must be implemented in $A$ so that $A$ and
    $C$ become identical \cite{part_dist}.  The  distance between $A$
    and $B$, the total distance, is then
    obtained in the same way as for $m_{\text{moved}}$, i.e., as the
    sum of the two partial distances.
\end{itemize}

These definitions give different results when applied to
general partitions. In order to understand how two partitions are
related to each other, both measures are useful, 
as shown by the example below.

\begin{example}[Distances between two partitions]
Let $n=9$, and let $A$ and $B$ be the two partitions
$A=\{ \{1,2,3,4,5,6\},\{7,8,9\} \}$ and
$B=\{ \{1,2,4,5,7,8\},\{3,6\},\{9\}  \}$.
The \textit{meet} from (\ref{meet}) is then given by
$C=\{  \{1,2,4,5\},\{3,6\},\{7,8\},\{9\}  \}$.
The measures computed according to the two  methods above are shown in
Table \ref{table:example_part_distance}.
It shows that the two definitions of
distance between partitions may give different measures. The
information provided by the two methods is significant, in fact both
the information on how many divisions must be performed or how many
elements must be moved are relevant in order to find out the
relationship occurring between partitions.
The distance from partition A to the meet is 3 according to the first
method, $m_{\text{moved}}$, and this
means that three elements (elements 3, 6 and 9) must be moved but we
don't know how these three
elements are grouped in partition B. This information is given by the
measure computed with the second method, $m_{\text{div}}$:
only two divisions are necessary and
this implies that ele\-ments 3,6 and 9 belong to two different clusters
in partition B.
When comparing partition B with respect to A,
the measure $m_{\text{moved}}$ gives the distance from partition B to the meet equal to
two, in fact elements 7 and 8 must be moved in order to obtain two
partitions with the same clusters. Furthermore, the measure $m_{\text{div}}$ says that
these two elements belong to the same cluster in A and therefore it
can be seen as a subcluster.
If the distance computed by $m_{\text{div}}$ is
much smaller than the distance given by $m_{\text{moved}}$, it means that many
subpartitions are present; elements belonging to the same original
partitions are grouped together.
\end{example}
\begin{table}[h]
  \begin{center}
    \caption{Distance between partitions A and B of Example 1 where
      $\text{distA}$ ($\text{distB}$)      
  denotes distance from A (B) to the meet and $\text{distAB}$ the
  total distance  between partitions.\label{table:example_part_distance}}
\begin{tabular}[h]{|c|c|c|c|}
\hline
 Method  &  $\text{distA}$ & $\text{distB}$ & $\text{distAB}$ \\ \hline
   $m_{\text{moved}}$    &    3              &       2          &   5          \\
   $m_{\text{div}}$    &    2              &       1          &   3          \\   \hline
\end{tabular}
\end{center}
\end{table}

For the partitions of Table \ref{table:modularity_karate} for the
Karate club, the distance
from the original division of the set (the one represented in
Fig.~\ref{fig:karate}) has been computed using both definitions
given above. The results are illustrated in
Table \ref{table:karate_distance}.

\begin{table}[h]
 \caption{Karate club: distance between the partitions with highest $Q$
   ($P$) and the meet,    and the original division ($P_0$) and the meet.}
   \centering
   \begin{tabular}{|l|c|c|c|c|c|}
\hline
Algorithm                      & No of clusters & \multicolumn{2}{|c|}{$m_{\text{moved}}$} & \multicolumn{2}{|c|}{$m_{\text{div}}$}\\
\cline{3-6}
                            &                    &  
                            $\text{dist}P$ &
                            ${\text{dist}P_0}$ &
                            $\text{dist}P$ &
                            ${\text{dist}P_0}$ \\
\hline \hline
Hierarchical (shortest path) &   4 &  0  & 11  & 0 & 2 \\
K-means (sum of paths)       &   2 &  0  & 0   & 0 & 0 \\
K-means (shortest path)      &   4 &  1  & 16  & 1 & 3 \\
\hline
    \end{tabular}
   \label{table:karate_distance}
  \end{table}

Table \ref{table:karate_distance} shows that the partition produced by
hierarchical clustering (shortest path), $P$,
represents a subpartition of the original division, $P_0$; in fact the
distance from the obtained partition $P$ to the meet is zero.
K-means algorithm (shortest path), instead, misclassifies one element: this can
be seen from the fact that the distance $m_{\text{moved}}$ from $P$ to the meet is one
meaning that one element must be moved in order to have the partition
$P$ equal to the original division. K-means algorithm with distance
given by the sum of all paths, clearly, detects a
partition that is identical to the original division occurred in the club.

\subsection{Similarity between partitions}

\begin{figure}[h]
  \centering
      \includegraphics[width=0.45\linewidth]{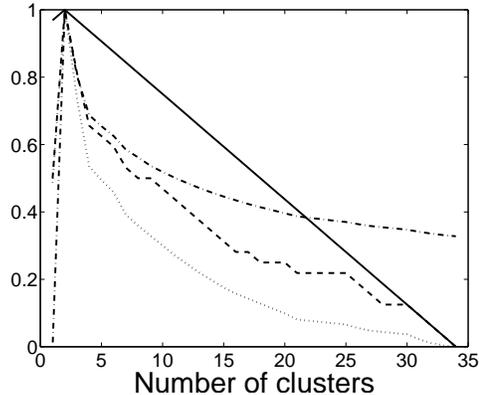}
   \caption{The four indices $I_{\text{div}}$ (full curve),
     $I_{\text{moved}}$ (dashed curve),
     $I_{\text{J}}$ (dotted curve) and
     $I_{\text{NMI}}$ (dashed-dotted curve) as function of number
     of clusters uncovered by hierarchical clustering (shortest path)
     for the Karate club.}
  \label{fig:indices}
\end{figure}

Two well-established indices for the similarity between two
partitionings of a set are the \emph{Jaccard index} and the
\emph{mutual information index}.
They are defined as:
\begin{itemize}
  \item[$I_{\text{J}}$:] The Jaccard index represents a measure of
    agreement between two partitions $A$ and $B$. It is defined
    as \cite{kuncheva2004}
    \begin{equation}
       I_{\text{J}}(A,B)=\frac{n_{11}}{n_{11}+n_{01}+n_{10}}
    \end{equation}
    where $n_{11}$ denotes the number of pairs of elements 
    that are simulataneously joined together in partition $A$ and $B$,
    $n_{01}$ ($n_{10}$) denotes the number of pairs of
    elements that are joined (separated) in $A$ and separated (joined) in $B$.
    It results in a matching coefficient in a range $[0 \; 1]$ where a
    value of 1 indicates that the two partitions are identical.
    The Jaccard index was originally developed to assess
    similarity among distributions of flora in different geographic
    areas \cite{jaccard1912}.
    
  \item[$I_{\text{NMI}}$:] The normalized mutual information index
    is a measure of similarity between partitions $A$ and
    $B$ \cite{kuncheva2004,strehl2002}. It is based on the mutual
    information between the partitions
    when the two partitions are treated as (nominal) random variables.
    The normalized mutual information index can be expressed as
    \begin{equation}
      I_{\text{NMI}}(A,B)=\frac{-2 \displaystyle{\sum_{i=1}^{|A|}} \displaystyle{\sum_{j=1}^{|B|}} n_{ij}^{ab}
        \log \left( \frac{n_{ij}^{ab} \cdot n}{n_i^a \cdot n_j^b} \right)}
      {\displaystyle{\sum_{i=1}^{|A|}}n_i^a\log\left( \frac{n_i^a}{n}\right)
      \displaystyle{\sum_{j=1}^{|B|}}n_j^b\log\left( \frac{n_j^b}{n}\right)},
    \end{equation}
    where $n_i^a$ represents the number of units in cluster $a_i$
    and $n_{ij}^{ab}$ denotes the number of shared elements 
    between clusters $a_i$ and $b_j$. It can be shown that $0\leq
    I_{\text{NMI}} \leq 1$ with $I_{\text{NMI}}(A,A)=1$.
\end{itemize}
In order to compare these two indices with the two distance measures introduced
in the previous section, we turn the measures into indices by
\begin{equation}
I_{\text{moved}}=1-\frac{m_{\text{moved}}}{n} \qquad \text{and} \qquad
I_{\text{div}}=1-\frac{m_{\text{div}}}{n},
\end{equation}
where $n$ is the number of units in the partioned set.
In Fig.~\ref{fig:indices} we show how these four indices vary for the
partitionings obtained by hierarchical clustering (shortest path) for
the Karate club, when we compare with the actual division which took
place.\\
The index $I_{\text{div}}$ is a straight line, which indicates that
one more node has been misclassified per  extra community which is
considered.
The other three indices behave somewhat similar, with the exception of
the mutual information, $I_\text{NMI}$, which does not tend to zero
for the limiting case when all nodes are placed in one cluster each.

As discussed in the previous section, the measures $m_\text{div}$ and
$m_\text{moved}$ can be interpreted as  the number of
divisions and the number of movements, respectively, which are
necessary for making two sets coincide.
Further, for $m_\text{div}$ and $m_\text{moved}$ we obtain also partial
measures which directly show if one partition is a subpartition of the
other.  
Because of this direct interpretation, the property of showing
subpartitions, and the similarity with the other measures when considered
as indices, we stick in the sequel of this paper to $m_\text{div}$ and
$m_\text{moved}$.

\section{Computer generated networks}
\label{section:CG}

Now we consider a class of computer generated random networks.
These networks, originally introduced in \cite{girvan02},
consist of 128 nodes each, divided into four communities
of equal size.
The links are distributed randomly, with the same probability for a
link to occur for each
pair of intra-community nodes, and another constant probability for
each pair of inter-community nodes, such that the average degree
of each node becomes 16.
Intra-community degrees, i.e., the part of the degree that
stems from links within the same community, are denoted as $z_{\text{in}}$,
and the inter-community degree as $z_{\text{out}}$.

\begin{figure}[h]
  \centering
      \includegraphics[width=0.85\linewidth]{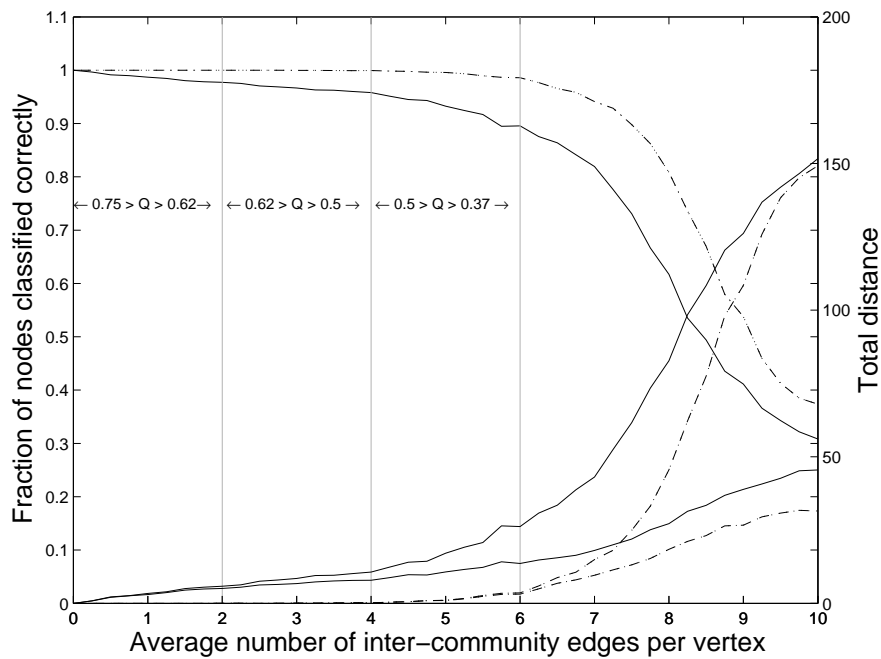}
   \caption{Fraction of correctly classified nodes (upper curves for
     low $z_{\text{out}}$) and distances from the true
     partition (lower curves for low $z_{\text{out}}$) for computer
     generated networks, as measured by $m_{\text{div}}$ and
     $m_{\text{moved}}$ ($m_{\text{div}} \le m_{\text{moved}}$).
     \textit{Solid curves}: hierarchical
    clustering (shortest path), \textit{Dashed-dotted curves}: 
    K-means clustering (shortest path and sum of paths coincide).
    Each point is an average over 100 different networks.
  Typical values of the modularity, obtained from the true partition,
  are marked.}
  \label{fig:correctlyclassified}
\end{figure}

By applying the algorithms introduced above (100 times) to the networks with
$0 \le z_{\text{out}} \le 10$, we can compare our methods to
some others studied in the literature.
In Fig.~\ref{fig:correctlyclassified} we show the fraction of
correctly classified nodes for our three algorithms for different
values of $z_{\text{out}}$.
It turns out that the K-means clusterings (both measures) give results which
are comparable to 
methods that are very recently developed in order to take care of this
community detection problem \cite{Duch-05}.
Indeed, the  results of the two K-means algorithms are identical,
which make these computer generated networks unique among the ones we
study.
This might reflect that these networks are in some sense less complex
than the real-world networks, but this issue deserves a further
investigation before any definitive conclusions can be drawn.
The fraction of nodes that are classified correctly by hierarchical
clustering (shortest path) is lower, but the
result  still makes sense as 60\% of nodes are correctly classified  when
$z_{\text{out}}=8$.\\

However, to measure only the fraction of correctly classified nodes
might give a wrong picture of performance, as noted also in \cite{Danon-05}.
For instance, if the community detecting algorithm happens to
divide one correctly identified cluster into two,
it is normally not a serious error, but still the fraction correctly
classified nodes will decrease drastically.

If we instead consider the measure $m_{\text{div}}$ introduced above, we
can see that the deviation from the real structure is quite
modest.
In Fig.~\ref{fig:correctlyclassified}, we also depict the
two distance measures $m_{\text{moved}}$ and $m_{\text{div}}$ (total distance),
which in some sense give
a better description of the outcome from the algorithms.
Here, however, we clearly see how these different descriptions give
similar results.


\section{College football}
\label{section:football}

United States college football is now considered
\cite{girvan02}. The network represents the schedule of
Division 1 games for the year 2000 season: the 115 vertices in the
graph represent
teams (identified by their college names) and the 613 edges represent
regular-season games between the two teams they connect. The community
structure is well-known. Teams, in fact, are divided into 13 conferences
containing around 8-12 teams each. Games are more frequent  between
members of the same conference than between members of different
conferences, with teams playing an average of about seven
intra-conference games and four inter-conference games in the year 2000
season.  Interconference play is not uniformly distributed, teams that
are geographically close to one another but belong to different
conferences are more likely to play one another than teams separated by
large geographic distances.

\begin{figure*}[h]
  \begin{tabular}{cc}
      \includegraphics[width=0.45\linewidth]{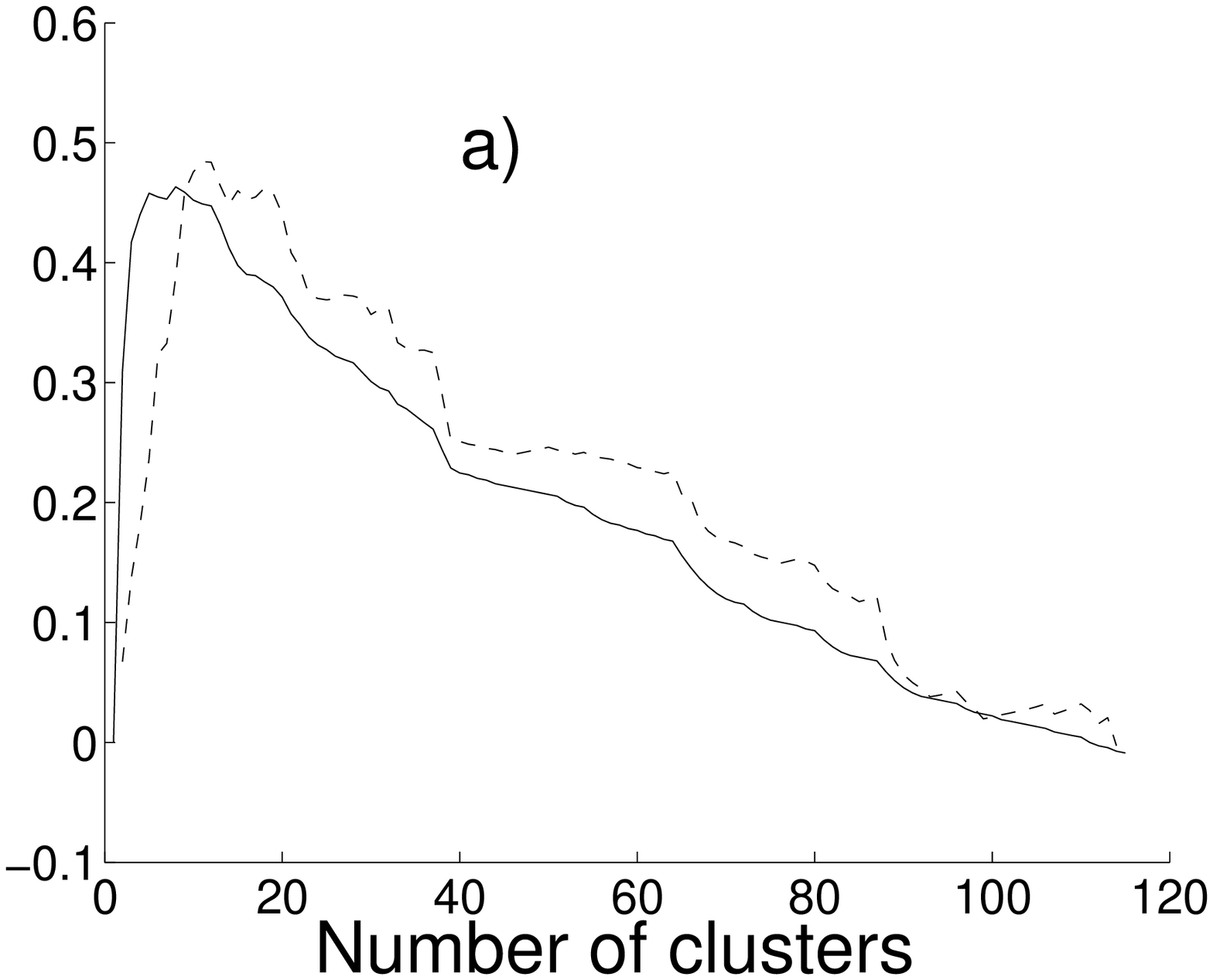}
      &
      \includegraphics[width=0.45\linewidth]{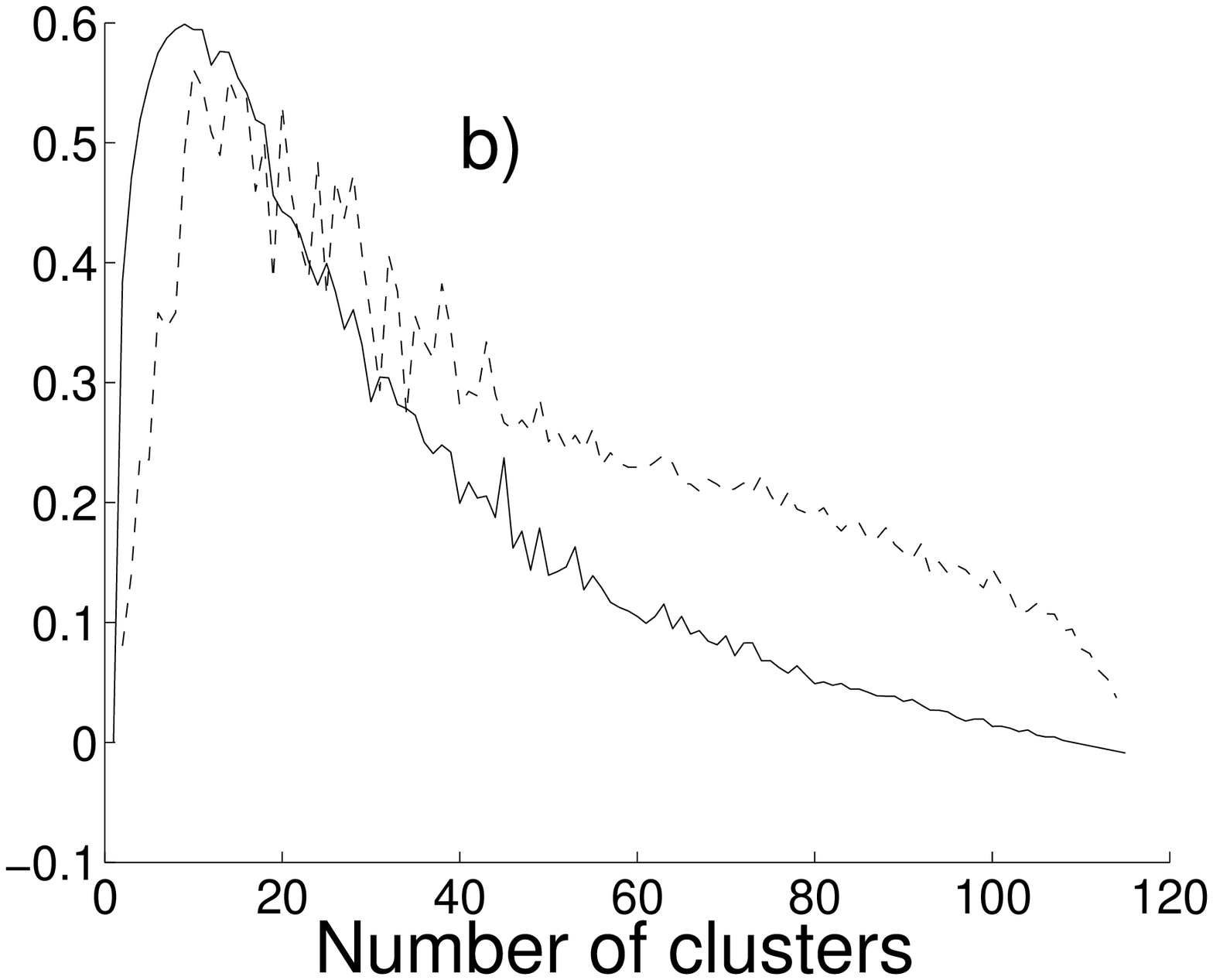}
      \\
      \includegraphics[width=0.45\linewidth]{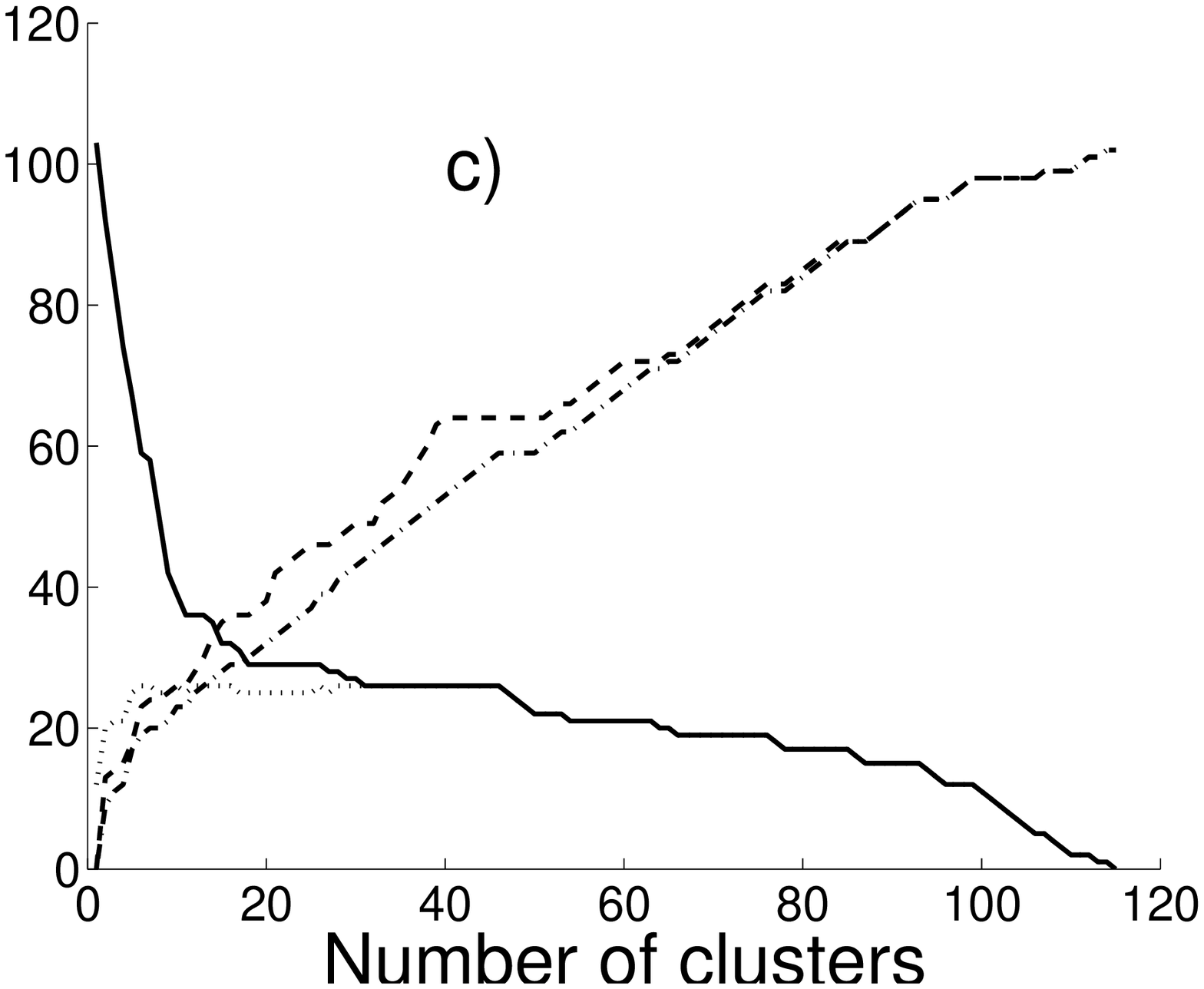}
      &
      \includegraphics[width=0.45\linewidth]{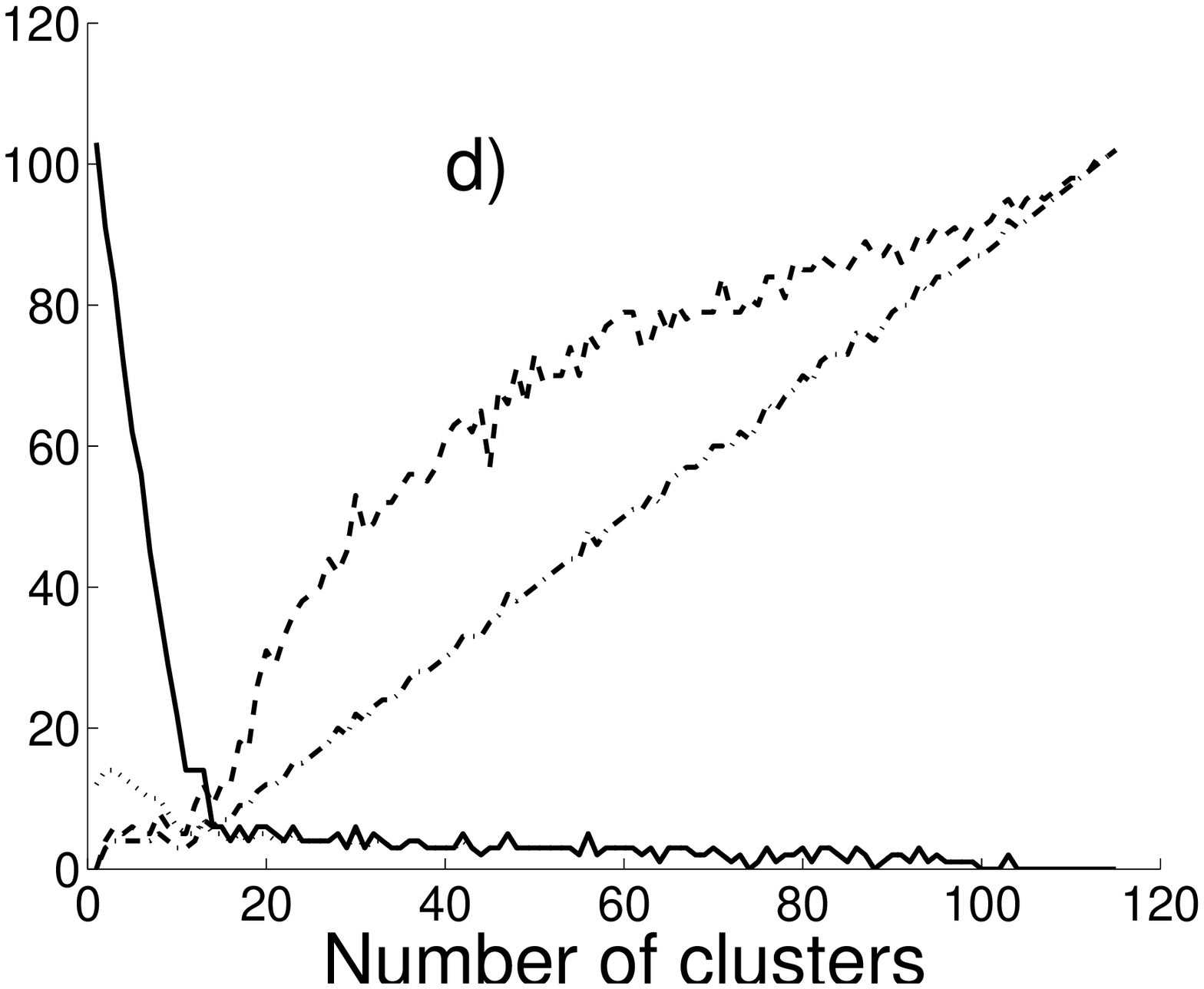}
    \end{tabular}    
    \caption{College football network. (a) Modularities $Q$ (solid
      curve) and
      coherence scores ($Z$-values) (dashed curve) depicted as functions of the number of clusters
      when the partition comes from hierarchical clustering (shortest
      path). (b) Same as (a) but for K-means (sum of paths). (c) Distances
      between the detected partition $P$ and the meet, and between the
      true partition $P_0$ and the meet, as
      functions of the number of clusters,
      when the partition comes from hierarchical clustering (shortest
      path). Solid curve: $m_{\text{moved}}(P,\,\text{meet})$, dashed
      curve: $m_{\text{moved}}(P_0,\,\text{meet})$, dotted curve:
      $m_{\text{div}}(P,\,\text{meet})$, dashed-dotted curve:
      $m_{\text{div}}(P_0,\,\text{meet})$.
      (d)  Same as (c) but for K-means (sum of paths).}
    \label{fig:football}
\end{figure*}

We have repeated the same clustering procedures as above
to this data and the results are reported in
Fig.~\ref{fig:football} and the most prominent features are emphasized
in Tables \ref{table:modularity_football} and
\ref{table:football_distance}.%
\footnote{The modularity for the
true division into conferences is $Q=0.537$, which  is
smaller than the peak we obtain, $Q=0.6044$.}
All algorithms are able to partition the network into
communities. 
The highest value of the modularity  $Q$ for each algorithm has
been determined and the result is reported in Table
\ref{table:modularity_football} together with the corresponding
community structure and the results of the null hypothesis test.
\begin{table}[h]
 \caption{College football network: highest modularities obtained using different
   approaches and their corresponding community structure computed for
   the true network ($Q$) and for the null hypothesis ($Q_{H_0}$).
   \label{table:modularity_football}}
  \centering
   \begin{tabular}{|l||c|c||c|c|}
\hline
Algorithm   & $Q$  & \#clusters & $Q_{H_0}$ & \#clusters \\
\hline \hline
Hierarchical  (shortest path)   & 0.46 & 8 & $0.23\pm 0.010$ &7\\
K-means      (sum of paths)    & 0.60 & 11 & $0.28\pm 0.006$ & 5\\
K-means      (shortest path)   & 0.60 & 10 & $0.27\pm 0.007$ & 5\\
\hline
    \end{tabular}
  \end{table}

The K-means algorithm gives similar results in both cases with distance between
nodes given by (\ref{sumpaths}) and given by the shortest path between
the nodes. Two different partitions are determined with the only
difference being that in the latter partition 
one conference is broken into two pieces
and grouped with 
two other conferences
This is the reason why we in Fig.~\ref{fig:football} (a) and (b) only show the
results from one of the K-means algorithms and from the hierarchical
clustering. 

The resulting partition is very close to the real conference structure
of the 2000 season. The fact that only 11 partitions are detected is
because two of the conferences are identified as one
conference (this happens also with the other algorithms).
There are 5 independent teams 
that do not belong to any conference and
they tend to be grouped with the teams with which they are most
closely associated. There are  only three teams that are
misclassified.

For the partitions of Table \ref{table:modularity_football} the distance
from the original division of the set has been computed using both
measures $m_{\text{moved}}$ and $m_{\text{div}}$.
The results are illustrated in Table \ref{table:football_distance}
and in Fig.~\ref{fig:football} (c) and (d).

\begin{table}[h]
 \caption{College football network: distance between the partitions (P) with
   highest $Q$ and the meet, and between  the original division
   ($P_0$) and the meet.
   \label{table:football_distance}}
   \centering
   \begin{tabular}{|l|c|c|c|c|c|}
\hline
Algorithm                      & No of clusters & \multicolumn{2}{|c|}{$m_{\text{moved}}$} & \multicolumn{2}{|c|}{$m_{\text{div}}$}\\
\cline{3-6}
                            &                    &  
                            $\text{dist}P$ &
                            $\text{dist}P_0$ &
                            $\text{dist}P$ &
                            $\text{dist}P_0$ \\
\hline \hline
Hierarchical (shortest path) &   8   &  50 & 24 & 25  & 20\\
K-means (sum of paths)       &   11  & 14  & 5  &  5  & 3 \\
K-means (shortest path)       &   10  & 21  & 9  & 9  & 6  \\
\hline
    \end{tabular}
  \end{table}
Here it is clear that the partition
closer to the original set is the one detected by K-means
algorithm (sum of paths) as distance between nodes, in fact
only 14 elements must be moved in order to make the set coincide with
the meet.

Both partitions obtained with K-means algorithm (using both definitions
of distance between nodes) correspond to high values of the modularity
 $Q$ so it can be interesting to see how \textit{distant} these
two partitions are between each other. Let $P_{K_W}$ denote the
partition obtained with the K-means algorithm (sum of paths) ($N=11$),
and $P_{K_G}$ the partition obtained with the K-means
algorithm (shortest path) ($N=10$).
The distances to the meet from respective partition become
$m_{\text{moved}}(P_{K_W},\,\text{meet})=6$,
$m_{\text{moved}}(P_{K_G},\,\text{meet})=12$,
$m_{\text{div}}(P_{K_W},\,\text{meet})=2$, and
$m_{\text{div}}(P_{K_G},\,\text{meet})=3$.
The total number of elements to be moved in the two partitions in
order for them to become identical is 18.
If 6 elements are moved
within the communities of  $P_{K_W}$, then  $P_{K_G}$ becomes a
subpartition (as the number of partitions is different).
These two partitionings are thus closer to each other than any of them
to the ``true'' partitioning, hence at least in this case the choice
of algorithm seems less crucial.

\subsection*{Coherence score}

If the true partition of the network is not known, we have to
turn to another way of validating the structure than to
measure the distance.
A common case is that we have annotations (one or many) to each node,
and that we should utilize in some way the property that some of these
annotations are common to more than one node.
Here we propose the following method,
illustrated by the simplest case
where each node has exactly one annotation:

Assume every node in the network has a classification---here it
is the conference to which it belongs.
To judge the quality of a specific community partitioning, let us
take one of the modules we detected as an example.
This module contains in total ten teams, with eight teams from 
one conference and
two teams from 
another conference.
Given the actual sizes of these two conferences, the total number
of teams and the total size of this module, we can from these
numbers calculate the probability that the eight
teams would come from the actual conference should have occurred by
chance (from the hypergeometrical distribution), and the same
probability for the  two teams from the other conference.\footnote{For reasons
  that will become clear in the next section, we do not calculate the total
  probability for this event.}
These probabilities are the  $p$-values for this instance, and
we assign the lowest one we find to this module.
For future processing, we store the negative logarithm of this
$p$-value.

However, it is problematic to consider these $p$-values as true probabilities
since there are many different tests we perform (one for each
class of nodes which exists within the actual module), i.e.,
this is a multiple testing problem.
It is likely that unlikely things should occur.
Therefore we pick eleven random teams and calculate the $p$-values
in the same way again, keeping the lowest one.
By repeating this many times, we obtain a distribution of
(negative logarithms of) $p$-values
which has a mean, $\bar{p}$, and a standard deviation, $\sigma_p$.
From this, we can construct standard $Z$-scores as
\begin{equation}
  Z=\frac{p-\bar{p}}{\sigma_p}
  \label{Z-score}
\end{equation}
These $Z$-scores indicate how unlikely the present distribution is,
such that the higher $Z$-score, the more improbable the actual
distribution, and the more relevance our found module has.
That is, they form a \emph{coherence score} for the community.

By considering all modules simultaneously, and proceeding in the
way described above, we obtain a global  coherence score for the
whole community structure.
In Fig.~\ref{fig:football}, we show the modularity, the
distances from the true partition (both $m_{\text{moved}}$ and $m_{\text{div}}$) and the
coherence-scores for different number of communities.
The algorithms employed are
hierarchical clustering (shortest path) in (a) and (c), and K-means (sum of
paths) in (b) and (d).
We clearly see how the coherence score and the modularity co-variate,
such that they essentially peak at the same number of
clusters. Remembering that the modularity is a graph-theoretical
measure relying only on network properties (paying no attention to
what the nodes represent) and that the coherence score does not
utilize that the underlying structure is a network but only focus on
the annotations of the nodes, this is a remarkable result.
Further, we see in (c) and (d) how the total distances, both
$m_{\text{moved}}$ and $m_{\text{div}}$, have minima approximately
where the modularity peaks. These observations support the tacit
assumption in many previous papers that the partition with the highest
modularity also corresponds to the one that makes most sense.

\section{Gene network of \emph{Saccharomyces cerevisiae}}
\label{section:gene_network}

As our final example, we consider a somewhat larger network where
no true partition exists (or at least is known). 
The network represents the  interactions between
regulatory proteins and genes in 
\textit{Saccharomyces cerevisiae} (ordinary yeast) \cite{milo2002}.
The 690 nodes represent genes,\footnote{A few are ``pseudogenes'', which
  here means that they are complexes, composed by two or more gene products.}
and the 1079 edges correspond to biochemical interactions.
The edges are directed from a gene that encodes for a transcription factor
to a gene transcriptionally regulated by that protein, i.e.,
we have a directed network.
However, here we remove all directionality from the edges in order to
more easily apply the same algorithms and measures  as before
in this paper.

\begin{figure*}[h]
  \begin{tabular}{cc}
          \includegraphics[width=0.45\linewidth]{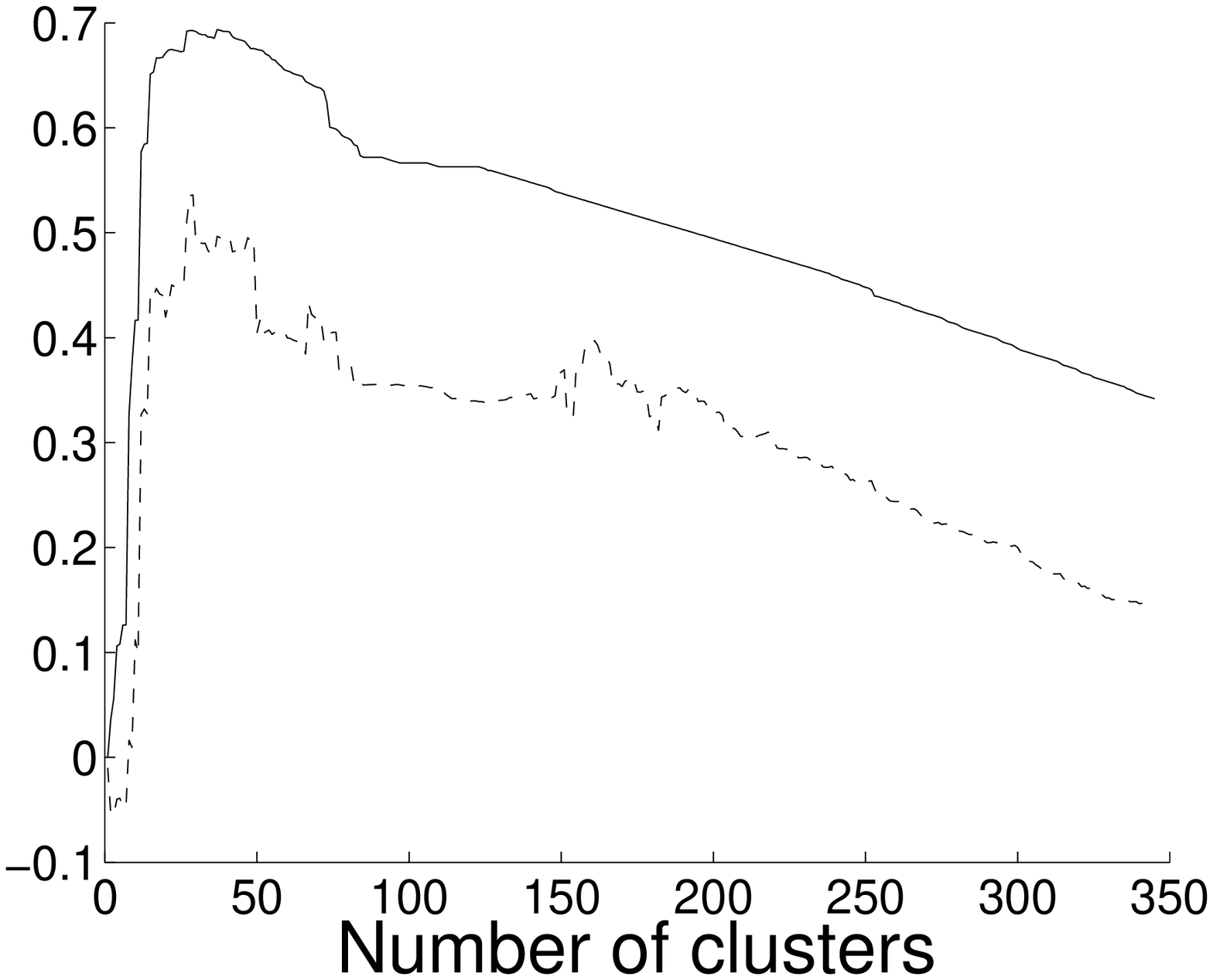}
      &
      \includegraphics[width=0.45\linewidth]{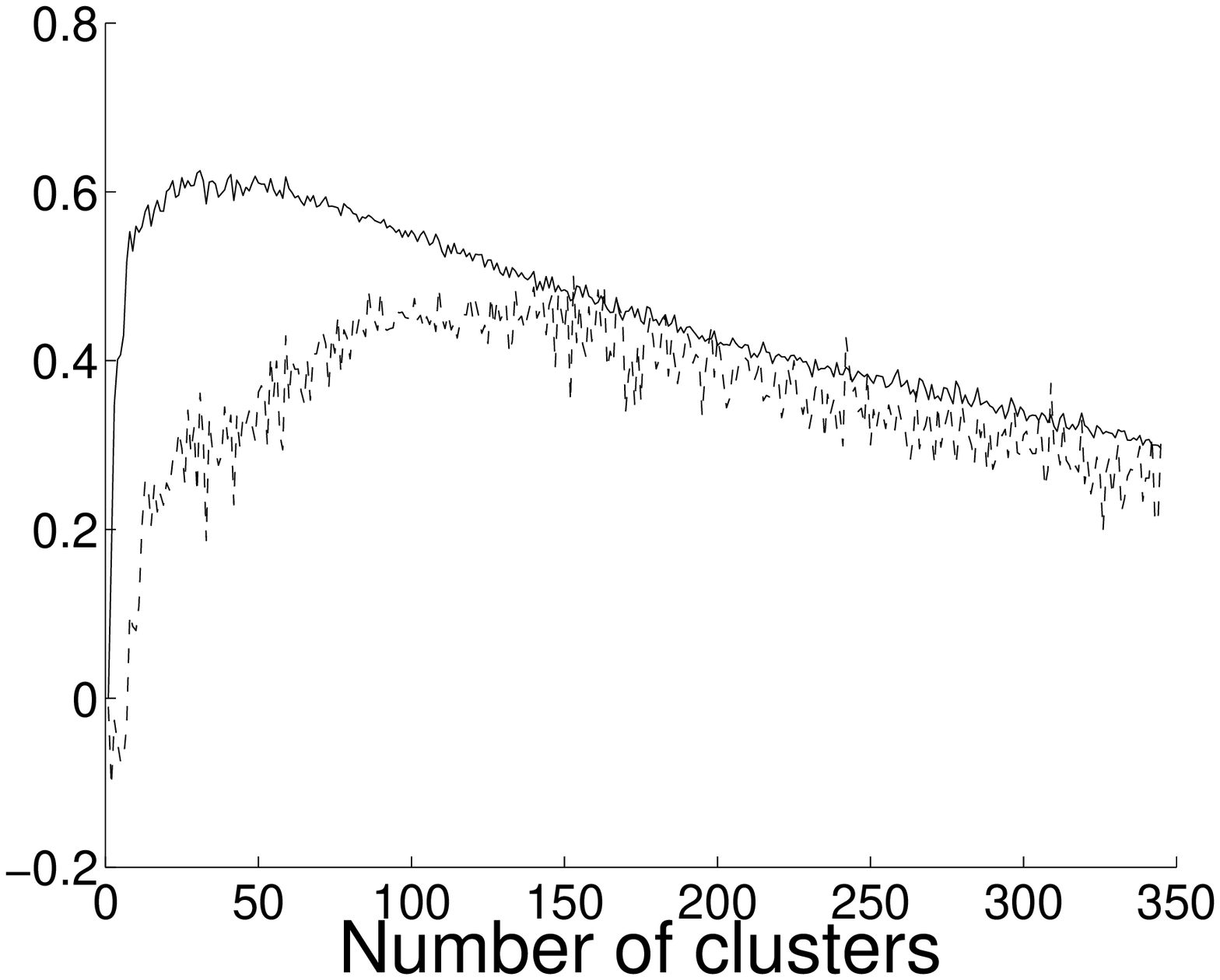}
    \end{tabular}    
  \caption{Modularity (solid curve) and coherence (dashed curve)
    for the communities detected in
    the gene network of \textit{Saccharomyces cerevisiae}. The
    coherence is scaled down with a factor of 100.
    Left: Hierarchical clustering (shortest path).
    Right: K-means clustering (sum of  paths).}
  \label{fig:yeast_plots}
\end{figure*}

We have partitioned this gene network
into several communities using the
algorithms described above.
For each number of communities, we have as before calculated
the modularity (\ref{modularity}) and the coherence (\ref{Z-score}),
see Fig.~\ref{fig:yeast_plots}.

\begin{figure}[h]
\centering\includegraphics[width=0.9\linewidth]{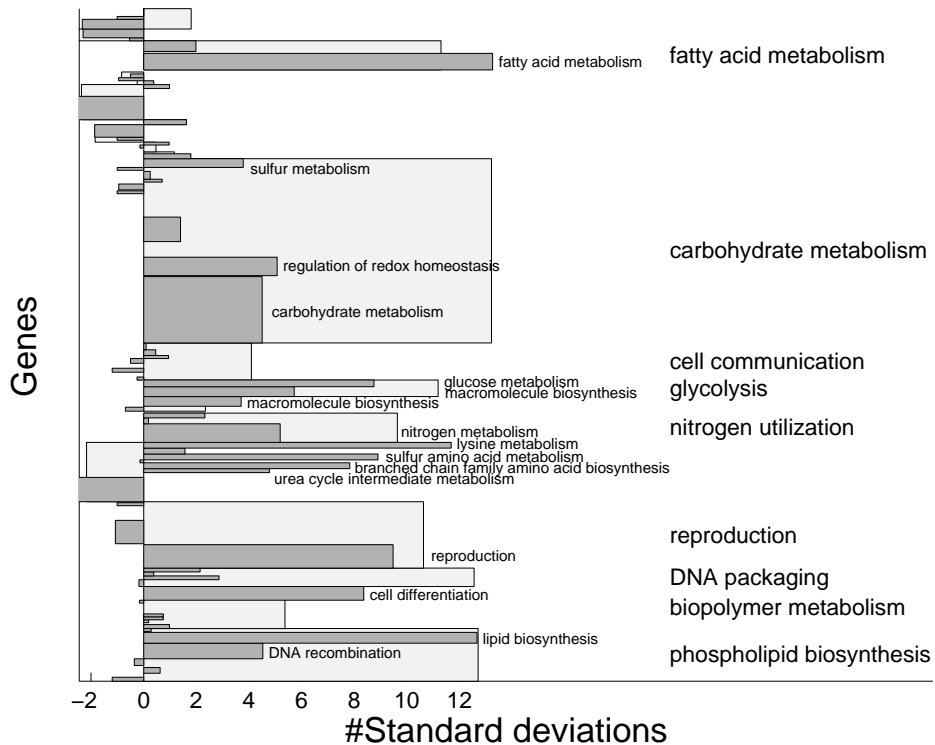}
   \caption{Coherence scores ($Z$-values) for the biological process terms,
     associated to each community. The light grey bars correspond to a
     division into 22 communities obtained with hierarchical
     clustering (shortest path) and has $Q=0.67$. The dark grey bars
     correspond to 177 communities, obtained with the same algorithm
     and with $Q=0.51$.
     The widths of the bars are proportional to the number of
     genes in each community.}
  \label{fig:Z-scores}
\end{figure}

The external validation for calculating the coherence is this time
somewhat less straightforward. 
We employ the Gene Ontology (GO) database \cite{GO} in order
to find a classification for the nodes/genes of the network.
In this database, the genes of \emph{S. cerevisae} (and several
other organisms) are arranged in a directed acyclic graph
according to which biological process they belong.\footnote{They can
  also be ordered according to biological function or cellular localization,  properties
  we do not utilize here.}
A gene is assigned both the ontology term for the process it belongs
to, as well as all terms for the 
parental processes in the graph, i.e., there are many different terms
associated with each gene.
To judge the quality of a specific community partitioning, we query
the database with a list of all genes in each community.
It is because of these multiple annotations of all genes we cannot
calculate one probability for the whole network division, as remarked
in the previous section.
These $p$-values are then treated the same way as for the football
teams, and we obtain coherence scores from the same kind of null
hypothesis as before.
In Fig.~\ref{fig:Z-scores} we show the distribution of all
 coherence scores for the divisions into 22 and 177 communities,
respectively,  obtained by hierarchical clustering (shortest path),
with corresponding 
$Q$-values of $Q=0.67$ and $Q=0.51$ (see Fig.~\ref{fig:yeast_plots}).

\begin{figure*}[h]
  \begin{tabular}{cc}
      \includegraphics[width=0.45\linewidth]{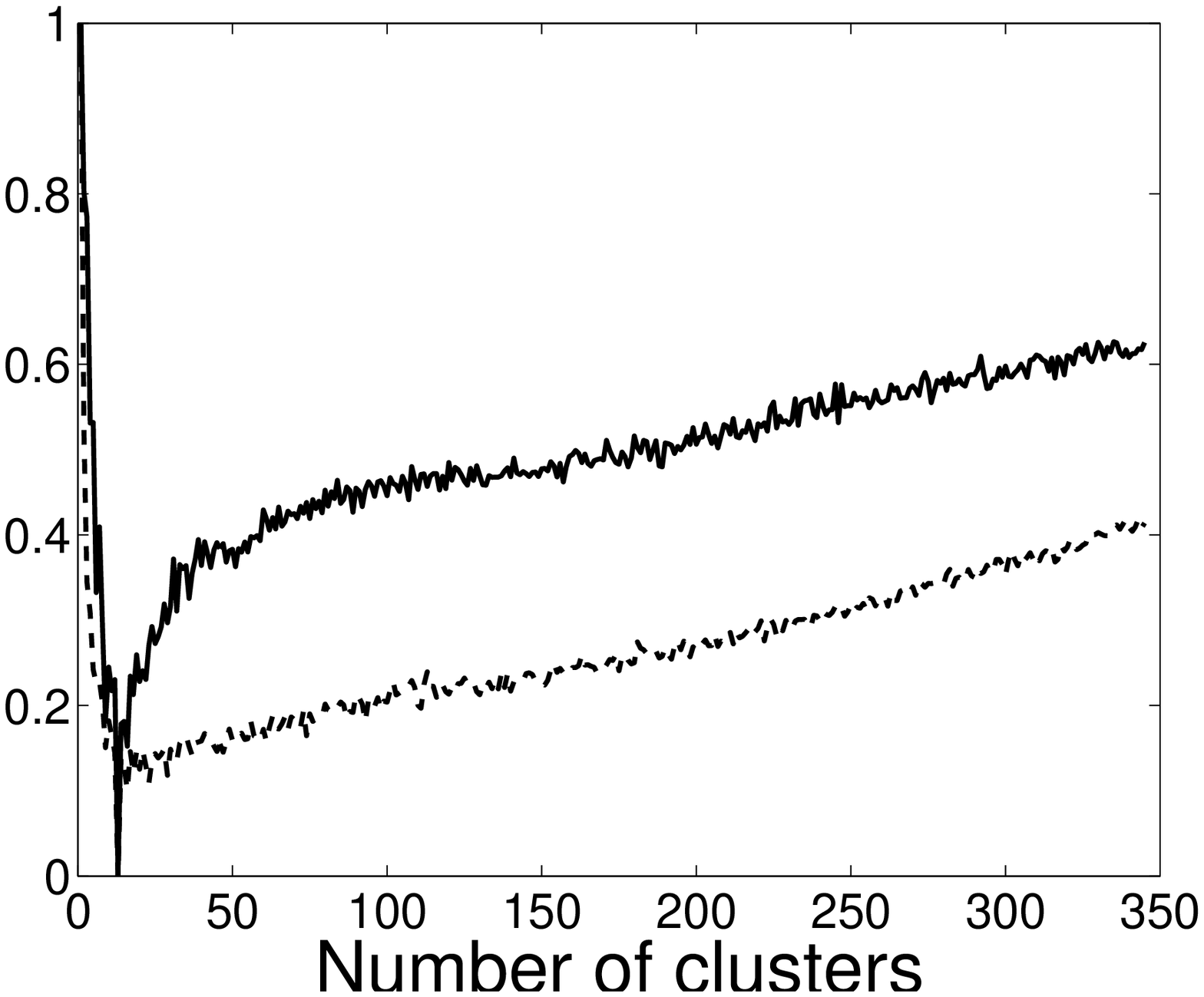}
      &
      \includegraphics[width=0.45\linewidth]{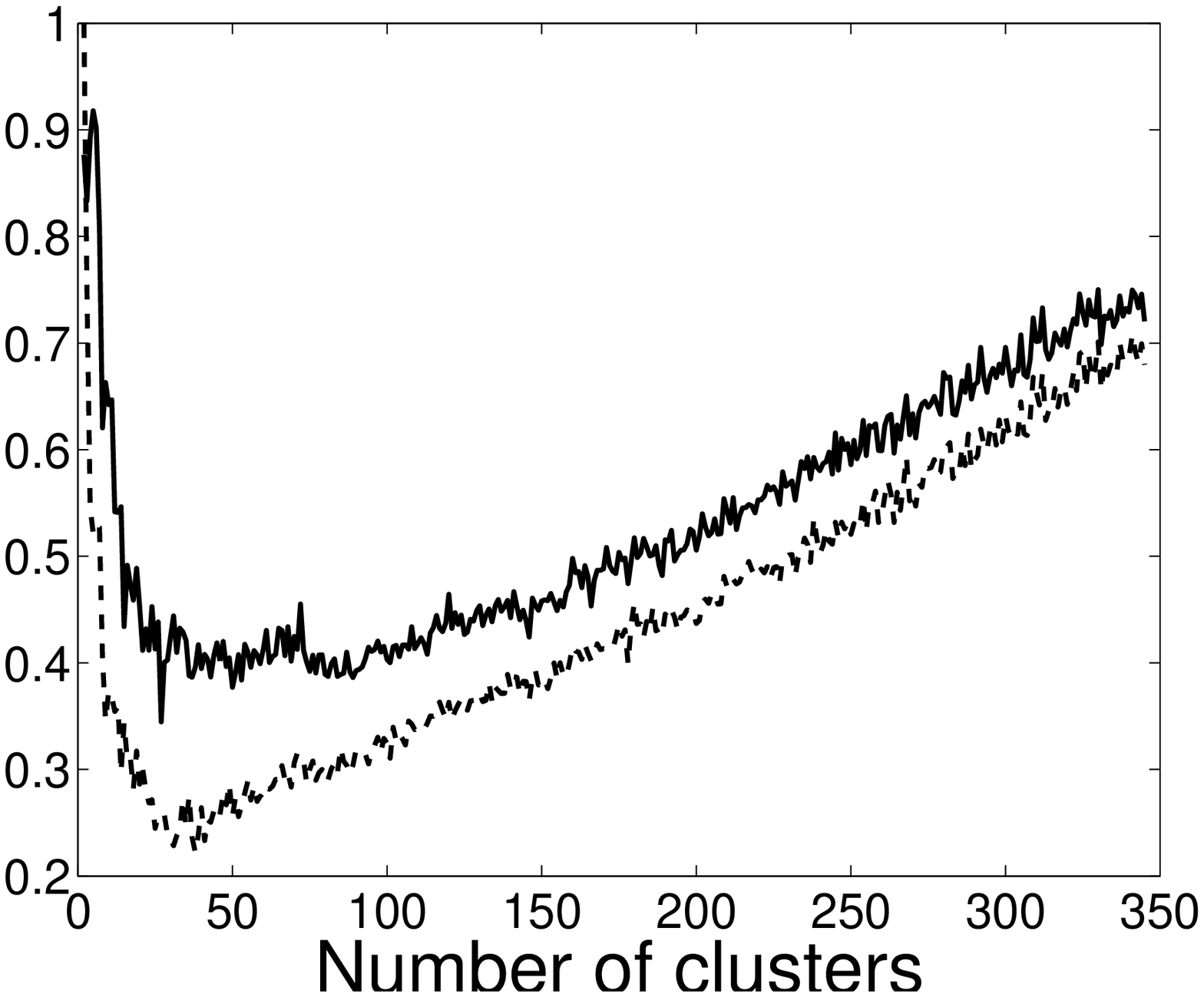}
    \end{tabular}   
   \caption{Normalized total distance between \emph{the} very best division
     obtained by K-means (shortest path) and K-means
     (shortest path) with a various number of communities (depicted
     at the $x$-axis) (left), and between partitionings obtained by
     hierarchical
     clustering (shortest path) and K-means (shortest path) for the
     same number of communities (right). Solid (upper) curves are
     $\hat{m}_{\text{moved}}$, while the dashed (lower) curves are $\hat{m}_{\text{div}}$.}
  \label{fig:gene_dist}
\end{figure*}

This time, we cannot measure the distance from any kind of ``true'' partition.
Instead we do a pairwise comparison between different methods.
In Fig.~\ref{fig:gene_dist} we show the normalized\footnote{Without
  prior knowledge of the network, it is hard to tell whether a
  distance is small or large just from the numbers $m_{\text{moved}}$
  and $m_{\text{div}}$. Therefore, we normalize these numbers by
  dividing them by the distances obtained as the mean of a sample of random
  partitionings with the same number of clusters and the same number
  of  units in each
  cluster as for the real network.} total
distances $\hat{m}_{\text{moved}}$ and $\hat{m}_{\text{div}}$ between
some different partitionings.
To the left, the figure shows the normalized distances between the
very ``best'' (highest modularity) division obtained with
K-means (shortest path) and the divisions obtained
by the same algorithm for the number of communities depicted along the
$x$-axis.
To the right, the figure shows the normalized distances between
the partitioning obtained by hierarchical clustering (shortest path)
and the one obtained by K-means (shortest path), for the number of
communities depicted along the $x$-axis.
Remarkably, we see how these distances have their minima not far from
where the modularity and the coherence score have their peaks.
By definition, the minima in the left panel have to be exactly zero, but worth
noting is how sharp these extrema are. The minima in the right panel are broader,
but still in a clear neighbourhood of the maxima of the modularity
and the coherence score. The conclusion is that the outcome of the
different clustering algorithms seem to coincide, as long as we are
satisfied with looking at the ``best'' partition (in the sense of
having the highest modularity). However, if we for some reason do not
strive for this optimum, the procedures can give rise to very
different results.

\section{Discussion and conclusions}
\label{section:discussionandconclusions}

We have in the present article presented various aspects of how to
find and evaluate community structures in complex networks.
The issue of finding such a structure is closely related to clustering
of data based on similarity over many experiments, without any
underlying network structure assumed.
This latter exploration is a standard tool today, and there are many
well-developed algorithms for handling the issue.
Here we show explicitly that at least some of these algorithms can be
used also for finding communities, indeed, one of them provided the
best partitioning in terms of high modularity we have found in the
literature thus far.
A drawback, however, is that these algorithms are computationally less
efficient.

Since there is no generally accepted definition of what really
constitutes a reasonable partitioning, we explored both Newman's
modularity and the so-called Silhouette index.
The former has become some kind of de-facto standard, and many authors
simply equate a proposed modular structure with having a high
modularity. As a contrast, we also investigated the Silhouette index,
but found rapidly that it was less suitable for networks.

The question of how distant two partitionings of the same network are
is somewhat new in the
present physics literature,\footnote{To the best of our knowledge, the only
exception is \cite{Danon-05}, which one of the referees draw our
attention to.} although the issue by no means is original.
We described two different such measures and discussed advantages and
drawbacks of each.  A combined use of both yields of course a better
description.
We also turned these to measure into indices and
compared them with the Jaccard index and the mutual information
index, showing that the behaviour was similar among those.

Finally, we proposed the introduction of a coherence score, indicating
the validity of the partitioning.
This score is especially well-suited when each node in  network
has several annotations, or belongs to annotations presented as a
directed acyclic graph, where children inherit all the parents'
annotations. This is the case for the Gene Ontology database, which
has become very popular recently in the bioinformatics community.
It turns out that the modularity and the coherence score peak for
approximately the same number of communities, at least for the
networks we have considered here.
Hence, this observation gives support to the standard procedure of
only striving for optimizing the modularity.

In summa, the issue of finding and evaluating community structures in
complex networks is an important part in the unraveling of properties
for the systems.
Here we have discussed various aspects of these themes, and also
introduced the new concept of ``coherence'' for a network.
We strongly believe this can be a valuable tool in the future for the
exploration of various networks.

\section*{Acknowledgment}
The authors thank M.~Girvan and M.E.J.~Newman for providing the data
of the college football network.
Financial support from CENIIT (Centre for Industrial IT at Linköping
University) and from the Carl Trygger's foundation is acknowledged.
\bibliographystyle{abbrv}
\bibliography{networks,bioappl,referenslista}

\end{document}